\documentclass{PoS}
\usepackage[utf8]{inputenc}
\usepackage{graphicx,epsfig,ragged2e}
\usepackage{amsmath,amssymb}

\title{MERGHERS: An SZ-selected cluster survey with MeerKAT}

\ShortTitle{An SZ-selected cluster survey with MeerKAT}

\author{\speaker{K.~Knowles}\thanks{Corresponding author.}, J.~Sievers\\
        Astrophysics \& Cosmology Research Unit, School of Chemistry \& Physics, University of KwaZulu-Natal, Durban 4041, South Africa\\ 
        E-mail: \email{kendaknowles.astro@gmail.com}}
        
\author{A.~Baker, J.P.~Hughes\\
        Department of Physics \& Astronomy, Rutgers, The State University of New Jersey, 136 Frelinghuysen Road, Piscataway, NJ 08854-8019, USA\\
        E-mail: \email{ajbaker@physics.rutgers.edu}}
        
\author{K.~Basu, M.W.~Sommer\\
        Argelander Institute for Astronomy, University of Bonn, Auf dem H\"{u}gel 71, 53121 Bonn, Germany\\
        E-mail: \email{kbasu@astro.uni-bonn.de}}        
        
\author{V.~Bharadwaj, M.~Hilton, K.~Moodley, S.P.~Sikhosana\\
        Astrophysics \& Cosmology Research Unit, School of Mathematics, Statistics \& Computer Science, University of KwaZulu-Natal, Durban 4041, South Africa\\
        E-mail: \email{moodleyk41@ukzn.ac.za}}  
        
\author{F.~de Gasperin, H.T.~Intema\\
        Leiden Observatory, Universiteit Leiden, PO Box 9513, NL-2300 RA Leiden, the Netherlands\\
        E-mail: \email{intema@strw.leidenuniv.nl}}
        
\author{R.~Deane, S.~Makhathini, O.~Smirnov\\
        Centre for Radio Astronomy Techniques and Technologies, Department of Physics and Electronics, Rhodes University\\
        E-mail: \email{o.smirnov@ru.ac.za}}        

\author{M.~Devlin, S.~Dicker, S.~Stanchfield\\
        Department of Physics \& Astronomy, University of Pennsylvania, 209 South 33rd Street, Philadelphia, PA 19104 USA\\
        E-mail: \email{devlin@physics.upenn.edu}}

\author{C.~Ferrari\\
        Laboratoire Lagrange, UMR 7293, Universit\'{e} de Nice Sophia-Antipolis, CNRS, Observatoire de la C\^{o}te d’Azur, 06300 Nice, France\\
        E-mail: \email{chiara.ferrari@oca.eu}}
        
\author{N.~Oozeer\\
        SKA South Africa, The Park, Park Road, Pinelands, Cape Town 7405, South Africa\\
        E-mail: \email{nadeem@ska.ac.za}}
               
\author{C.~Pfrommer\\
        Leibniz-Institut für Astrophysik Potsdam (AIP), An der Sternwarte 16, 14482 Potsdam, Germany\\
        E-mail: \email{cpfrommer@aip.de}}
        
\author{K.~van der Heyden \\
        Astrophysics, Cosmology \& Gravity Centre, Department of Astronomy, University of Cape Town, Private Bag X3, Rondebosch 7701, South Africa\\
        E-mail: \email{heyden@ast.uct.ac.za}}
        
\author{J.~Zwart\\
        Department of Physics \& Astronomy, University of the Western Cape, Private Bag X17, Bellville, Cape Town 7535, South Africa\\
        E-mail: \email{jzwart@uwc.ac.za}}
        
\abstract{The MeerKAT telescope will be one of the most sensitive radio arrays in the pre-SKA era. Here we discuss a low-frequency SZ-selected cluster survey with MeerKAT, the MeerKAT Extended Relics, Giant Halos, and Extragalactic Radio Sources (MERGHERS) survey. The primary goal of this survey is to detect faint signatures of diffuse cluster emission, specifically radio halos and relics. SZ-selected cluster samples offer a homogeneous, mass-limited set of targets out to higher redshift than X-ray samples. MeerKAT is sensitive enough to detect diffuse radio emission at the faint levels expected in low-mass and high-redshift clusters, thereby enabling radio halo and relic formation theories to be tested with a larger statistical sample over a significantly expanded phase space. Complementary multiwavelength follow-up observations will provide a more complete picture of any clusters found to host diffuse emission, thereby enhancing the scientific return of the MERGHERS survey.}

\FullConference{MeerKAT Science: On the Pathway to the SKA,\\
	25-27 May, 2016,\\
		Stellenbosch, South Africa}

\begin{document}

\section{Introduction}

%
%

Galaxy clusters, being at the intersection of astrophysics and cosmology, are a rich class of objects to study throughout the electromagnetic spectrum. They typically consist of $\sim$70-80\% dark matter, and 15-20\% of a tenuous plasma called the intracluster medium (ICM), with the remainder of the mass budget taken up by the baryonic matter of the member galaxies. 

To date, there exist large samples of cluster detections through both X-ray and microwave (through the Sunyaev-Zel'dovich (SZ) effect \cite{SZ1972}) observations \cite{Ebeling2001, Bohringer2004, LloydDavies2011, Vanderlinde2010, Menanteau2013, Planck2014}. Using these large statistical samples, cosmological parameters such as $\sigma_8$ and the matter density, $\Omega_m$, can be constrained via cluster number counts and/or mass functions. However, since the cluster mass is inferred from observed cluster properties such as X-ray luminosity or the SZ Compton-$y$ parameter, number counts and mass functions are sensitive to the chosen mass-observable scaling relation. These scaling relations are influenced by the cluster physics and dynamical state, and understanding the thermodynamic properties of the clusters in the sample can consequently reduce the systematic uncertainties in the scaling relations and in the cosmological parameter constraints. One method for determining a cluster's dynamical properties is through the detection of diffuse cluster-scale ($\sim$Mpc) synchrotron emission in the form of radio halos and relics, the existence of which have been linked to merger activity in the cluster.

Radio relics are polarized structures found on the cluster outskirts and are typically elongated with the major axis perpendicular to the cluster radius. They have so far been found only in disturbed systems and have a physical extent larger than expected based on the typical life-time of the emitting particles. These properties indicate a merger connection, with the particles re-energised through diffuse shock acceleration (DSA) caused by merger shocks. However, some radio relics challenge our understanding of the physics involved: relics have been found with Mach numbers which are incompatible with current reacceleration/DSA formation theories \cite{Macario2011}, some clusters with strong X-ray shocks have no relic emission \cite{Russell2011}, and the alignment of magnetic fields within some relics is still not well understood \cite{vanWeeren2010}.

Radio halos are centrally located, $\sim$Mpc scale regions of diffuse synchrotron emission, with typical flux densities of a few $\mu$Jy/arcsec$^2$, fainter than their relic counterparts. To date they have been found in approximately 50 high mass ($M_{500} > 4.7 \times 10^{14} M_\odot$) clusters, most of which lie at low to intermediate redshifts ($z < 0.4$), with only a handful detected above a redshift of 0.5 (see \cite{Yuan2015} and references therein\footnote{Yuan et al. (2015) contains the most recent catalogue of radio halo, relic, and mini-halo observations.}). As in the case of radio relics, the observed size of radio halos establishes the requirement for particle reacceleration. Two main theories have been studied: the hadronic or secondary-electron model in which relativistic electrons are created from proton-proton collisions within the ICM \cite{Dennison1980, BlasiCola1999}, and the primary electron or turbulent reacceleration model which stipulates that an existing population of cosmic ray electrons are reaccelerated via merger-driven turbulence \cite{Ensslin1998, Brunetti2004}. Although a hybridization \cite{BrunnettiLazarian2011} of the two models has not been ruled out, the latter model is the currently preferred theory due to its predictions of a bimodality in the radio-X-ray plane and the existence of a population of ultra-steep spectrum sources ($\alpha \sim 1.5 - 1.9$), both of which have been observed (see e.g. \cite{Brunetti2007}, \cite{Bonafede2012}).

The aforementioned bimodality separates the radio-X-ray plane into radio halo-loud clusters whose non-thermal and thermal properties are correlated, and a population of radio halo-quiet systems with upper limits approximately an order of magnitude below the correlation. The ultra-steep spectrum sources mostly lie slightly below the scaling relation, potentially filling in the region between the correlation and the upper limits \cite{Cassano2013}. Initially discovered with X-ray clusters, the bimodality has also been found using a sample of SZ-selected clusters, although there is some evidence that the bimodality isn't as strong in this case \cite{Cassano2013, SommerBasu2014}, with the radio halo dropout fraction being smaller for the SZ versus X-ray selection \cite{SommerBasu2014}.

In line with the primary electron model predictions, magnetohydrodynamical simulations by Donnert et al. \cite{Donnert2013} showed radio halos to be transient phenomena, with the strength of the radio emission dependent on the stage of the host cluster merger. Such simulations have been compared to radio halo observations, along with multiwavelength data, to estimate where along the merger track a radio halo is being observed \cite{Knowles2016}. This transitory characteristic may be a key underlying factor in the uncertainties in global radio halo properties and the scatter within the various scaling relations.

In addition to the impact of selection effects on the bimodality, there are still a number of open questions related to radio halo physics and formation, chief among them being a full understanding of the formation model and merger connection {\textendash} radio halos have been found in low-luminosity clusters \cite{Giacintucci2005}, as well as in a cluster with an intact cool core \cite{Bonafede2014}. The powerful radio halo (and relics) found in the high redshift cluster ``El Gordo'' \cite{Lindner2014} also questions our understanding of the magnetic fields driving the synchrotron emission: inverse Compton losses are expected to dominate at high redshift, significantly reducing the observed radio power in these systems.

A third form of diffuse radio emission in clusters are radio mini-halos (see e.g. \cite{vanWeeren2014}). These are associated with the brightest cluster galaxy (BCG) in cool-core systems and are not connected with cluster merger activity. There are only 20 or so mini-halos found to date as it can be difficult to disentangle their faint, diffuse emission from that of the BCG itself. Although the physics and formation of mini-halos are not fully understood, it has been suggested that they form from gas sloshing in the centre of the cluster and that there may be a formation link with their larger radio halo counterparts. Furthermore, there is evidence for a spatial correspondence between mini-halo emission and high pressure regions seen in high resolution SZ observations \cite{Ferrari2011}, linking them to the non-thermal ICM. 

For all cases of diffuse radio emission found in clusters, large statistical samples of the different types of emission are necessary in order to answer the open questions relating to their global properties, as well their connection to cluster dynamics. To this end, a large survey of a few hundred clusters with a sensitive instrument such as MeerKAT would be highly beneficial to start addressing some of these questions in the pre-SKA era.

In this proceeding we will discuss the possibility of such a study with the MeerKAT telescope, the MeerKAT Extended Relics, Giant Halos, and Extragalactic Radio Sources (MERGHERS) survey, as a MeerKAT open time project. In Section \ref{sec:clustersample} we discuss a potential cluster sample, and highlight the strengths of MeerKAT for diffuse emission studies in Section \ref{sec:mkatstrength}, with some technical issues addressed in Section \ref{sec:sims}. Sections \ref{sec:descience} and \ref{sec:otherradioscience} describe the types of radio-based science one can achieve with a large statistical sample, and the multiwavelength follow-up is discussed in Section \ref{sec:mwscience}. We conclude and summarise the proceedings in Section \ref{sec:summary}.

\section{Cluster sample}
\label{sec:clustersample}
Most of the radio halo, relic and mini-halo studies to date have been based on X-ray selected cluster samples. With the success of various SZ telescopes such as the Atacama Cosmology Telescope (ACT, \cite{Swetz2011}), the South Pole Telescope (SPT, \cite{Ruhl2004}), and Planck \cite{Planck2011}, we now have access to large SZ cluster samples which can be used for diffuse emission studies. SZ-selection has the benefit of the flux limit translating directly into a mass limit, since the integrated pressure scales with mass and is not diluted with redshift. There is also some indication that the SZ may select more uniformly in merger timescale based on the fact that the signal boosting during a merger is less severe for SZ than in X-rays \cite{Poole2007}. 

To expand the discovery space of diffuse emission studies to low mass and high redshift, we need a large, homogeneously selected cluster sample covering a wide range of mass and redshift. This is ideally provided by ACT which is mass-limited, with better sensitivity and resolution than the Planck satellite. Planck beam dilution is also severe at higher redshift, which compromises the translation between flux-limited and mass-limited in this phase space. 

ACT has been online since 2007, having completed three seasons with its polarization-capable upgrade, ACTPol, in 2015. ACTPol is already producing $\sim$100 clusters in the first 680 deg$^2$ (Hilton et al., in prep.), thereby probing low-mass clusters over a much wider redshift range than Planck. ACTPol covers 9 hours in Right Ascension broken into two strips: 23.5 - 2.5 hr at $\delta$ = -2.5, and 10 - 16 hr at $\delta$ = +7.5 and overlaps with the Northern Baryon Oscillation Spectroscopic Survey \cite{Schlegel2007}. ACTPol has improved sensitivity to low-redshift systems compared to its predecessor survey, and once the full 2700 deg$^2$ of ACTPol coverage has been analysed, we expect to have a sample of a few hundred clusters over a wide range of redshift up to $z \sim 1$. The successor to ACTPol, Advanced ACT (AdvACT, \cite{deBernardis2016, Henderson2016}), will survey a total of 20,000 deg$^2$, which will significantly expand the cluster sample. 

The ACTPol sample, with a potential extension to AdvACT, will provide a large, homogeneously selected, statistical sample. The size of the sample allows for a split in redshift and/or mass to study the evolution of cluster radio emission, and would be expected to discover new forms of diffuse emission that don't fit within the three groups mentioned above (radio halos, relics, and mini-halos).

\section{Observing clusters with MeerKAT}
\label{sec:mkatstrength}
\begin{figure}
 \centering
 \includegraphics[height=4.3cm,clip,trim=35 0 60 40]{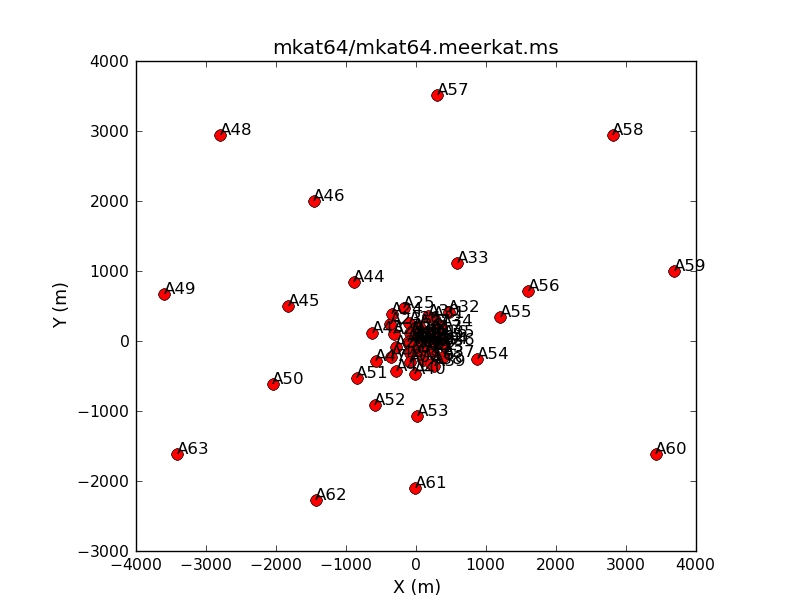}
 \includegraphics[height=4.5cm,clip,trim=2 0 0 3]{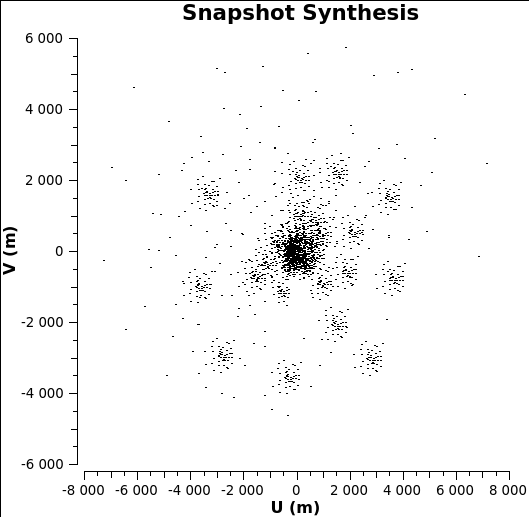}
 \includegraphics[height=4.5cm,clip,trim=2 0 0 3]{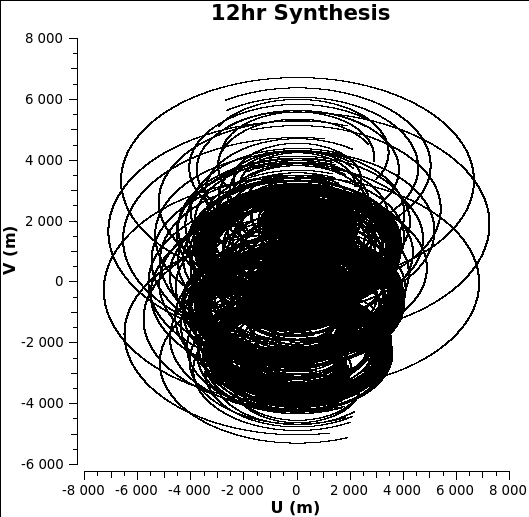}
 \caption{{\textit{Left:}} Layout for the full 64-dish MeerKAT array. {\textit{Middle:}} Snapshot uv-coverage for a source close to zenith. {\textit{Right:}} uv-coverage from a 12 hour synthesis on the same source.}
 \label{fig:ants_uv}
\end{figure}

The key requirements for successful interferometric observations of faint, diffuse cluster radio emission is flux sensitivity to extended structures and high enough resolution to disentangle point source emission from any fainter extended structures. The former is provided by short baselines and the latter by long baselines, both of which are simultaneously available with the full 64-dish MeerKAT array, thus making it highly suitable for diffuse emission studies. The array configuration and uv-coverage for a source close to zenith are shown in Figure \ref{fig:ants_uv}. We use the UHF (580 - 1015 MHz) band for our calculations as, due to their steep spectral indices, radio halos and relics are brighter at lower frequencies. This can be split into lower and upper sub-bands of approximately 200 MHz each, with central frequencies of 680 MHz and 890 MHz, respectively.

MeerKAT has a dense core made up of 70\% of the dishes. The minimum baseline is 29 metres. At the lower part of the MeerKAT UHF band, this corresponds to a maximum scale of 52 arcminutes to which the array is sensitive. Since diffuse cluster emission is expected to have an angular scale of 1 - 5 arcminutes, the dense core ensures that extended flux will not be resolved out. The array configuration also has an outer component, the longest baseline of which is 8 km. At 680 MHz, this corresponds to a theoretical minimum resolution of 15 arcseconds, which is small enough to distinguish between large scale emission and any compact point sources embedded therein up to a redshift of $z \approx 0.7$. 
 
\begin{figure}
 \centering
 \includegraphics[width=0.7\textwidth,clip,trim=0 0 0 23]{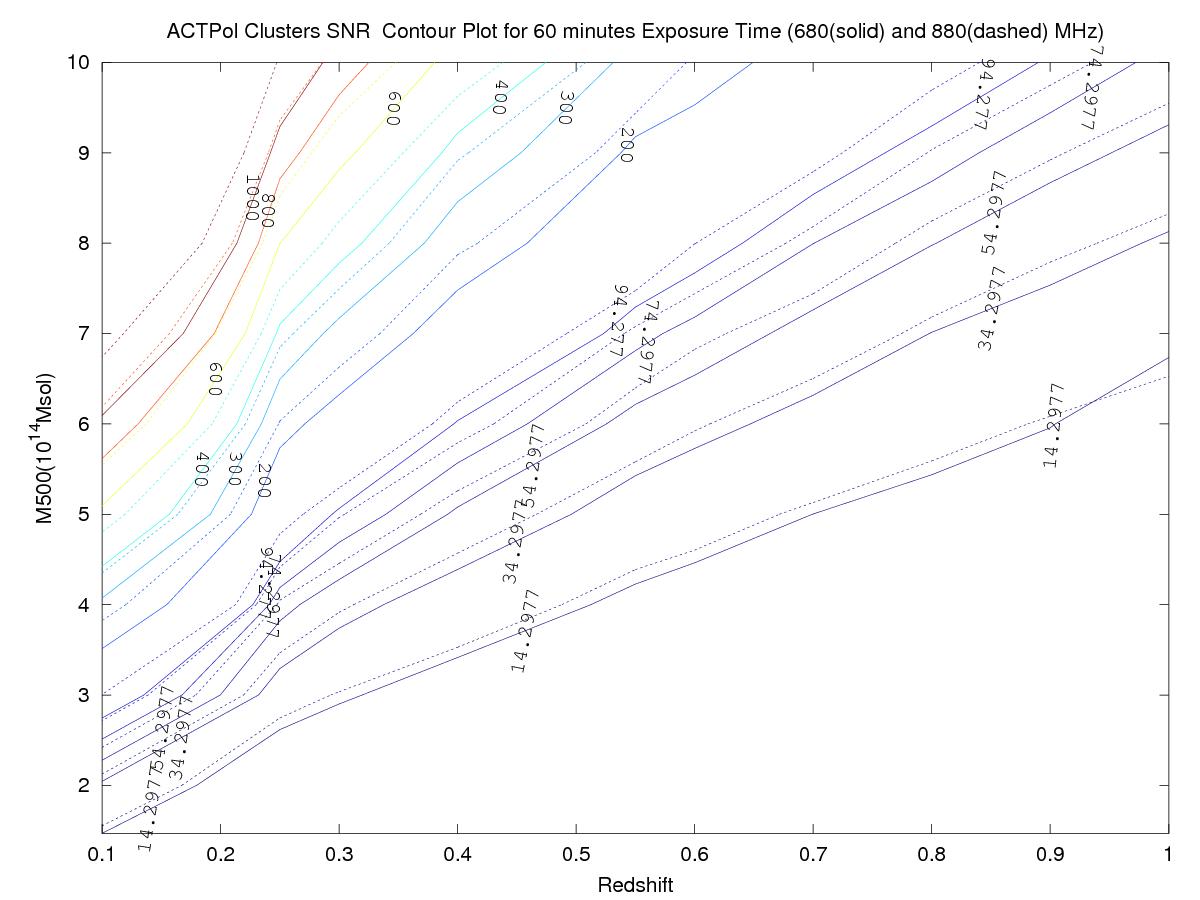}
 \caption{Recovered signal-to-noise of a simulated radio halo in test clusters with varying mass and redshift. The simulation is for one hour of on-source observing per UHF sub-band. The central sub-band frequencies are 680 MHz (solid) and 890 MHz (dashed).}
 \label{fig:sensitivity}
\end{figure}

Furthermore, MeerKAT's superior sensitivity, particularly in the short baselines, significantly reduces the integration time required to reveal any diffuse emission. A large statistical cluster sample can therefore be observed in a reasonable time of a few hundred hours, with approximately 1 - 2 hours on source per target. Simulations testing the sensitivity of MeerKAT to an extended, faint source show that a typical radio halo in a low-mass, high-redshift ($M_{500} = 4 \times 10^{14} M_\odot, \; z = 0.5$) cluster can be detected with a signal-to-noise of 15 per UHF sub-band in one hour of target observing, as seen in Figure \ref{fig:sensitivity}. Since the entire UHF band will be used, the sub-bands can be used for spectral index studies without the need for observing in an additional band.

\section{Technical considerations}
\label{sec:sims}
\begin{figure}
 \centering
  \hspace*{-0.1cm}
 \includegraphics[width=5.3cm,clip,trim=35 20 40 40]{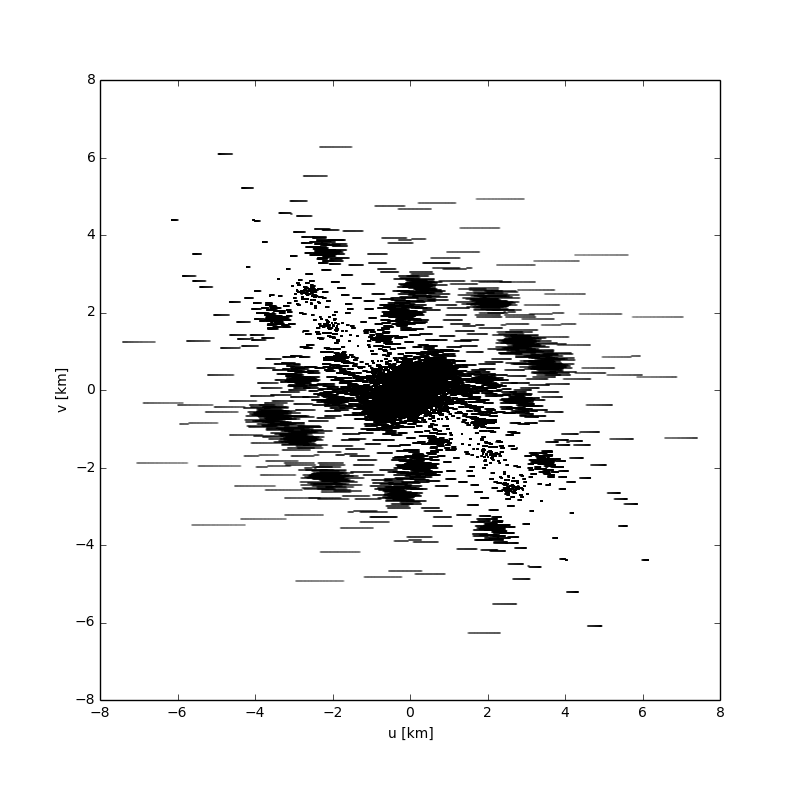} 
 \hspace*{0.65cm}
 \includegraphics[width=5.3cm,clip,trim=35 20 40 40]{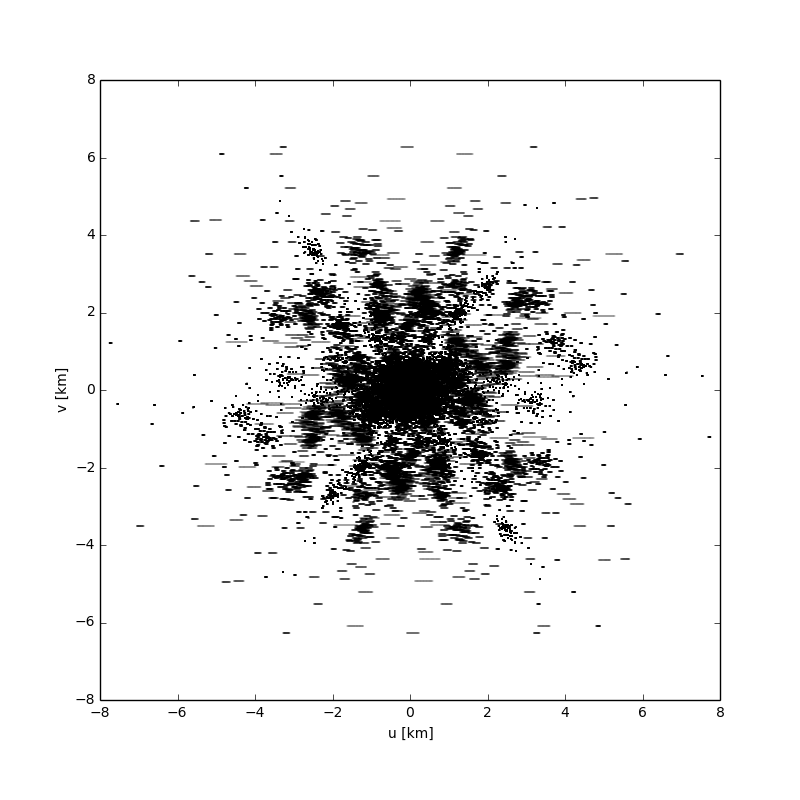}\\
 \hspace*{-0.5cm}
 \includegraphics[height=5cm]{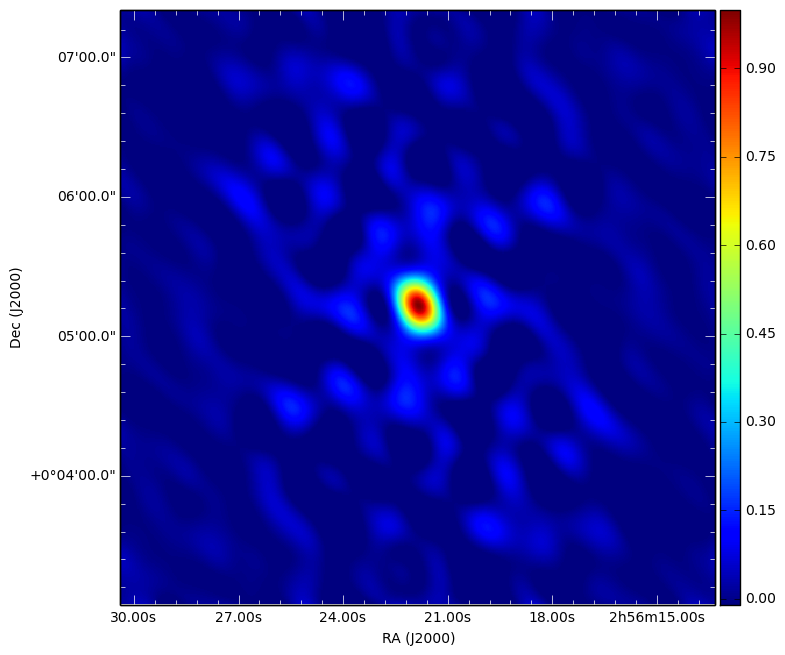}
 \hspace*{0.1cm}
 \includegraphics[height=5cm,clip,trim=5 65 40 75]{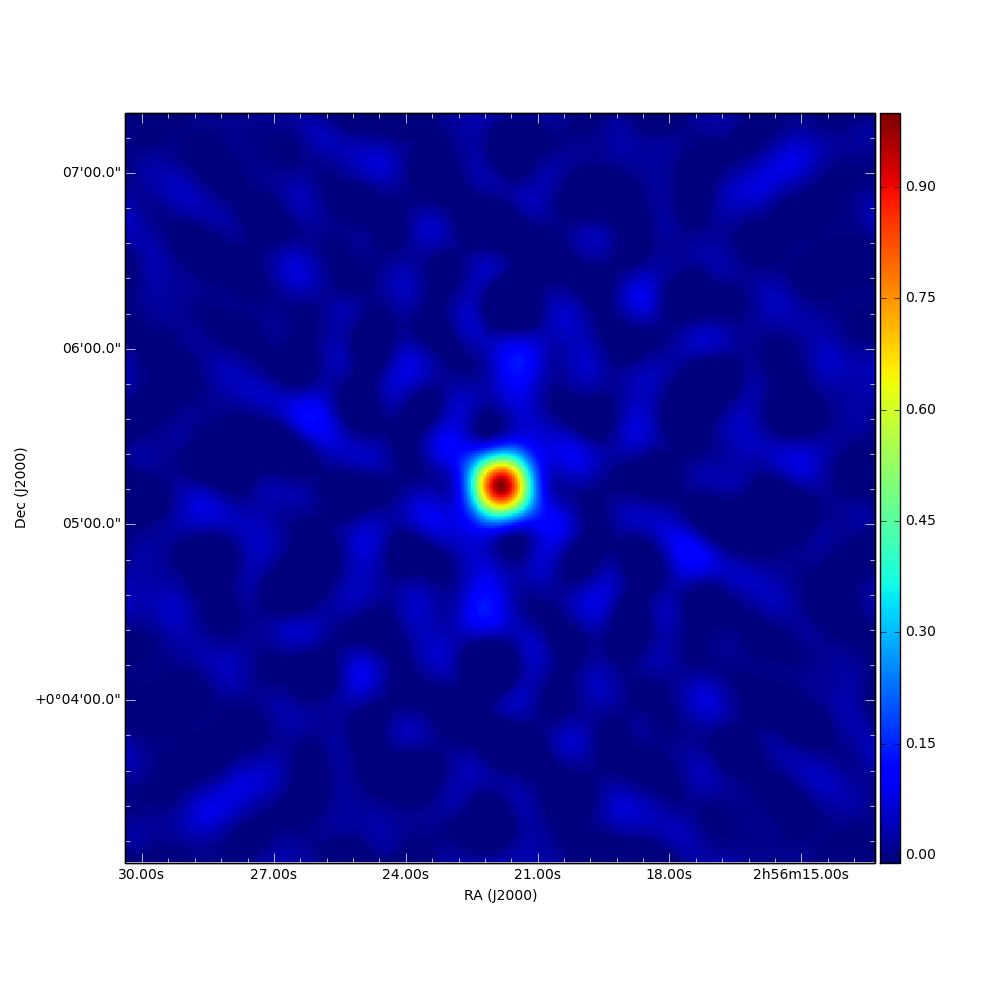}
 \caption{Simulations of the uv-coverage (top) and PSF (bottom) for one hour of target observation. {\textit{Left panels:}} One continuous 1-hour observing block. {\textit{Right panels:}} Three 20-minute observing blocks evenly spaced over a 6 hour period.}
 \label{fig:sims}
\end{figure}

Although one of the strengths of MeerKAT is its excellent sensitivity, the small amount of observing time required per target can lead to poor uv-coverage which hinders the reconstruction of a good image. One of the ways to circumvent this issue is to split an observation into shorter blocks spread over hour angle. Figure \ref{fig:sims} shows the uv-coverage (top panels) and point spread function (PSF, bottom panels), for two simulations each an hour long. Observation A (left panels) is one continuous observation block, and observation B (right panels) has been broken up into three blocks of twenty minutes, with a spacing of $\sim$2 hours between each block. Although both uv-coverages are patchy, observation B has more uniform coverage than observation A, with the result that its PSF has fewer sidelobes and the shape of the synthesised beam is more circular. This solution to the problem of poor uv-coverage is also well-suited to including regular polarization calibrator observations. In order to correctly calibrate the target polarization, good parallactic angle coverage is required for the polarization calibrator, which means it is necessary to regularly observe the relevant calibrator source throughout the target observation.

Another problem which has to be considered is that of classical confusion, which may become a hindrance to achieving the theoretical noise limits based on the array properties. Classical confusion occurs when faint sources smaller than the beam blend together, creating a natural noise limit which cannot be improved upon by longer integration times. The number of source solid angles, above a flux density $S$ is given by
\begin{equation}
 \beta = \left[ \left( \int_S^\infty n(x) dx \right) \Omega_s \right]^{-1}
\end{equation}
\noindent where the integral defines the source count above a given flux density, and $\Omega_s$ is the source solid angle, which can be written in terms of the source size $\theta_s$ as $\Omega_s = \pi \theta_s^2 [4 \ln 2]^{-1}$ \cite{Condon2009}. The flux density below which one is affected by confusion noise is generally calculated when $\beta = 25$ for power-law source counts. We estimate that the lower part of the UHF band is confusion limited at $\sim$10 $\mu$Jy. This noise limit can be achieved with approximately a 1 hour observation, indicating that images may become confusion-limited quite quickly. We note, however, that Bayesian data reduction techniques can take classical confusion noise into account, improving the accuracy of the source detection results and mitigating the negative effects (e.g. \cite{Feroz2009}).


\section{Prospective radio science}
\label{sec:radioscience}
\subsection{Diffuse cluster emission}
\label{sec:descience}
A targeted survey of a few hundred homogeneously selected clusters will allow for a statistical study of radio halos, relics, and radio mini-halos. In such a sample, based on existing statistics, one may expect to find a few score mergers with radio halo and/or relic emission, and a mini-halo in a few dozen of the relaxed clusters. As we move to the unexplored regions of the discovery space, i.e. low masses and high redshifts, we expect to find several systems with diffuse emission which cannot be classified by any of the existing categories. These systems will be extremely interesting to study in terms of whether they represent a transitional state between one or more of the known structures, adding to our understanding of how these structures form and evolve. 

With a statistical sample of radio halo detections, a primary goal would be to test the turbulent reacceleration model and to investigate the observed bimodality. Since current models and observations are restricted mainly to high-mass, low-redshift clusters, the expanded discovery space enables a test of the existence of the bimodality over a wider range of mass and redshift, as well as a study of the evolution of the scaling relations. Moreover, with a statistical sample, stacking of marginal or non-detections becomes viable. Radio halos in SZ-selected ACT clusters have already been detected \cite{Knowles2016, Lindner2014}, with more detections expected in an ongoing campaign to follow-up ACT clusters with the GMRT (Knowles et al. in prep). 

Much of what is known about the physics of radio relic formation is based on in-depth studies of a few spectacular examples \cite{vanWeeren2011, vanWeeren2012}. A large sample of quality relic detections possible with MeerKAT, based on a homogeneously selected cluster sample, will enable a study of global relic properties for the first time, as well as whether previously well-studied relics are representative of the general population. With observations probing lower mass systems where shock energies are expected to be lower, as well as higher redshift systems, a greater understanding of the physics and circumstances required to generate relic emission can be achieved. Moreover, the sensitivity and large field of view of MeerKAT may reveal the presence of relic emission not coincident with target clusters, which can be used to identify new clusters (e.g. \cite{Macario2014}).

Since the existing group of clusters studied in terms of diffuse emission is fairly inhomogeneous, a proper study of the drop-out fractions and non-detections for both halos and relics has not be possible due to varying selection effects throughout the population. With a large homogeneously selected sample, the non-detections can be statistically studied to constrain the life-cycle of these transient phenomena. In particular, with a large homogeneous sample including both relaxed and disturbed clusters, the evolutionary relationship between radio halos and mini-halos, if any, can be investigated. 


Extending the phase space of studied clusters will allow us to investigate whether magnetic fields scale in a self-similar way as one moves to lower mass and/or higher redshift systems. The evolution and scaling of magnetic fields with cluster properties would impact our understanding of both radio relic and radio halo formation. The wider research space will also allow for an investigation into the impact of merger properties on the formation of large-scale diffuse emission.

\subsection{Other radio science}
\label{sec:otherradioscience}
Although the primary science goal of the cluster observations is the detection and study of diffuse cluster emission, MeerKAT will provide useful, sensitive data for several other science objectives.

With the full polarization capabilities and excellent sensitivity of MeerKAT, the magnetic field in the clusters can be studied through Faraday rotation of background sources (e.g. \cite{Vacca2012}) or by polarized bent tailed radio sources within the cluster itself (e.g. \cite{EilekOwen2002}). Radio galaxies in many clusters show the presence of tailed emission. These objects are called narrow-angle or wide-angle tailed galaxies (NATs and WATs). The physics of the bent jets are still widely studied, with their morphologies thought to be influenced by environmental effects within the ICM, such as ram pressure or buoyancy forces \cite{Burns1998}. With the large dataset provided by a few hundred cluster observations, many NATs and WATs are expected to be found. In the higher redshift systems where better resolution is required to investigate the jets, these observations can serve as a basis for higher resolution radio follow-up. For those systems hosting bent tailed sources, they can be used to confirm the cluster dynamical state \cite{PfrommerJones2011}, assist in modelling merger dynamics \cite{Dehgan2014}, and measuring density and velocity flows within the ICM \cite{Freeland2008}. 

With a significant sample of relaxed cluster observations over a wide range of redshift and mass, the radio data will provide an excellent dataset for use in BCG studies, AGN feedback, and other compact radio source research: recent studies have shown that a large fraction of BCGs in relaxed, cool-core clusters are radio-loud AGN with non-thermal synchrotron jets and lobes. These sources create cavities in the thermal ICM (see e.g. \cite{Gitti2010}), and the study of the interaction between AGN lobes and the X-ray-emitting plasma is crucial to gaining a full understanding of BCG properties and the physics in the central region of galaxy clusters.

Finally, due to the large MeerKAT field of view in the UHF band ($> 1.5$ deg$^2$), superb sensitivity, and targeting a new sample of clusters, many previously undiscovered sources will be found in the full field of view, such as diffuse emission from nearby galaxies, or extended radio emission from known sources, which have been too faint to observe with other telescopes.


\section{Multiwavelength programme}
\label{sec:mwscience}
MeerKAT observations of a large statistical sample of galaxy clusters will provide a wealth of radio information for both diffuse emission science as well as many other cluster and galaxy related studies. The results of the radio observations can serve as a basis for multiwavelength follow-up on a variety of instruments and wavebands. 

Optical spectroscopy from a telescope such as SALT would provide spectroscopic redshifts for cluster member galaxies which are crucial for radio halo merger and timescale analyses (e.g. \cite{Knowles2016}), as well as providing an independent determination of the cluster dynamical state. For studies of NATs and WATs, optical spectroscopy would provide accurate galaxy redshifts to be used in modelling the merger activity in the cluster. With optical spectra, one can also study star formation of cluster members as a function of redshift and cluster dynamical state (see e.g. \cite{Ferrari2006} and references therein).

High-resolution SZ (MUSTANG-2) and X-ray (Chandra/XMM) imaging of the clusters hosting diffuse emission would probe both the non-thermal and thermal regions of the ICM, and give independent results on the cluster dynamical state. The results of the multiwavelength imaging are useful for modelling the merger and for timescale analyses of radio halos, and for determination of merger shock properties necessary for radio relic studies. For those clusters hosting mini-halos, high-resolution SZ imaging would provide more information on the global properties of mini-halos and their link to non-thermal pressure regions \cite{Ferrari2011}. In terms of AGN studies, high-resolution SZ observations can be used to image radio plasma bubbles and related features linked with AGN feedback, while X-ray imaging is necessary to probe the link between AGN feedback and X-ray cavities.

Finally, higher resolution radio observations with the JVLA or VLBI would be crucial for a full investigation into any observed AGN jets or bent tailed sources, as the low frequency MeerKAT beam cannot resolve the small-scale features necessary to fully analyse the jet morphologies. In addition, wide-field VLBI observations of a subsample of the clusters allows for a discrimination between the star forming (SF) and AGN nature of a cluster member's radio emission \cite{Herrera2016}, and therefore an investigation of the AGN/SF properties as a function of cluster radius. Given the power of the SZ selection, such observations would enable a systematic comparison of the AGN/SF properties as a function of cluster mass, redshift, and merger state.

\section{Summary and conclusion}
\label{sec:summary}
Galaxy clusters host a range of interesting astrophysical processes and their non-thermal properties can be studied through the presence of diffuse radio emission in the form of radio halos, relics, and mini-halos. High-quality observations of these large-scale radio signatures inform our understanding of plasma physics in terms of shocks and turbulence in the ICM, and act as proxies to study the properties of galaxy clusters and their dynamics. Under the assumption of a strong merger connection, observing diffuse radio emission in a cosmological cluster sample can assist in understanding the dynamical state of the cluster and its impact on cosmological constraints derived from the cluster mass function.

In order to significantly advance the study of these sources, a large, homogeneously selected, statistical sample needs to be observed. The SZ-selected ACTPol sample provides a few hundred clusters over a wide range of both mass and redshift. 

In order to achieve high-quality detections, a very sensitive instrument is required. MeerKAT is well-suited to this task due to its array configuration and superb flux sensitivity, which allows for the observation of a few hundred clusters within a reasonable time frame. With MeerKAT providing a sensitive radio dataset for the cluster sample, a multiwavelength follow-up of a subset of the sample that hosts diffuse emission will enhance the study of diffuse radio emission and other radio sources.

\end{document}